\documentclass[aps,showpacs,12pt,amsfonts]{revtex4}

\begin{document}
\title{Fermion Zero Modes in Odd Dimensions}
\author{Hyunsoo Min}\email{hsmin@dirac.uos.ac.kr}
\affiliation{Department of Physics, University of Seoul, Seoul 130-743, Korea}


\begin{abstract}
We study the zero modes of the Abelian Dirac operator in any odd dimension. We use the stereographic projection between a $(2n-1)$ dimensional space and a $(2n-1)$ sphere embedded in a $2n$ dimensional space. It is shown that  the Dirac operator with a gauge field of uniform field strengths in $S^{2n-1}$ has  symmetries of SU($n$)$\times$U($1$) which is a subgroup of SO($2n$). Using group representation theory,  we obtain the number of fermion zero modes, as well as their explicit forms, in a simple way. 
\end{abstract}
\pacs{11.10.Kk, 11.27.+d}

\maketitle

\section{Introduction}
The fermionic zero modes of the Dirac operators in a gauge theory are important in many areas of quantum field theory and mathematical physics. They play key roles in understanding axial vector anomalies and related topological properties\cite{current}. For example, soliton solutions carrying a fermionic number\cite{jackiw} can be generated by the fermionic zero modes in a semiclassical approximation. 
These zero modes have a deep relation with the 
massless fermions in Kaluza-Klein theory (see e.g., \cite{kk}). 
In even dimensions,  there is a powerful index theorem\cite{atiya} which describes a well-known relation between the zero modes and the topology of gauge fields and we have a rich spectrum of physical contexts studied. In odd dimensions, we have a different situation. There exists Callias index theorem\cite{callias} but the relevant topology in this case is given by the Higgs fields.

In three dimension, Loss and Yau\cite{loss} found the zero modes of the Abelian Dirac operator (without Higgs field) while studying the stability of hydrogen-like atoms with a sufficiently high nuclear charge in ultra-strong magnetic fields.  In \cite{adam,adam2}, the degeneracy of zero modes in this system was identified, and its possible relation with the topological index of the Hopf map was also studied. In plasma physics, a magnetic field carrying a magnetic helicity (which corresponds to the Hopf index) was experimentally observed\cite{plasma}. The possibility of topological electromagnetic radiation with magnetic and electric helicities is discussed in \cite{ranada}.

Recently, it has shown\cite{dunne} that such zero modes can easily be constructed if one uses the stereographic projection between $\textbf{R}^3$ and $S^3$. This construction may be extended to higher dimensions in a straightforward manner.  In this work,  we use this projection to identify 
the fermionic zero modes and conditions for their existence in arbitrary odd dimensions. 
First, we show that the symmetry group of the Dirac operator in $S^3$ is SU($n$)$\times$U($1$), a subgroup of SO(2$n$). It then follows that the dimension of a suitable irreducible representation determines the number of zero modes. Using the spherical harmonics in $2n$ dimensions, we also identify explicit forms of these zero modes.

\section{The stereographic projection}
In this section, we introduce the stereographic projection from $\textbf{R}^{2n}\supset S^{2n-1} \to \textbf{R}^{2n-1}$ as used in \cite{dunne}. 
We denote the coordinates $x_\mu$ ($\mu=1,\cdots, 2n-1$) in an arbitrary odd dimensional space  $\textbf{R}^{2n-1}$ and the coordinates $y_a$ ($a=1,\dots , 2n$) in the associated even dimensional space $\textbf{R}^{2n}$. The restricted space $S^{2n-1}$ is obtained by imposing the condition $y_a^2=1$. 
We may then define the stereographic projection from $S^{2n-1}$ to $\textbf{R}^{2n-1}$ via
\begin{eqnarray}
y_\mu=\frac{2x_\mu}{1+\vec{x}^2} \quad , \quad y_{2n}=\frac{1-\vec{x}^2}{1+\vec{x}^2},
\label{stereo}
\end{eqnarray}
with $\vec{x}^2=x_\mu^2$.

Consider a $2n$-dimensional gauge field corresponding to a constant field strength, 
\begin{eqnarray}
{\mathcal A}_a&=&-\frac{\mathcal F}{2}\left(y_2, -y_1, \dots ,y_{2n}, -y_{2n-1}\right)
\nonumber\\
&\equiv& -\frac{\mathcal F}{2} C_{ab}\, y_b ,
\label{sd-gauge}
\end{eqnarray}
with the antisymmetric matrix $C={\rm diag}(i\sigma_2, \dots, i\sigma_2, i\sigma_2)$. 
The matrix $C$ determines the structure of the field strength tensor as $F_{ab}={\cal F}C_{ab}$. 
We may take a different sign for each $i\sigma_2$ term in $C$, so that it corresponds to a different parity convention. For instance, the sign-flip in the last element corresponds to the choice of Loss and Yau when $n=2$. We may assume that ${\cal F}>0$ without loss of 
generality. The gauge field in (\ref{sd-gauge}) satisfies the Fock-Schwinger gauge condition: $y_a{\mathcal A}_a=0$. 

Analogous to the projection  $S^4\to\textbf{R}^4$ in \cite{adler,jackiwrebbi,bogomolny}, the projected $(2n-1)$ dimensional gauge field $A_\mu$ has the form: 
\begin{eqnarray}
A_\mu=\frac{\partial y_a}{\partial x_\mu}{\mathcal A}_a.
\label{projected-gauge}
\end{eqnarray}
A simple computation leads us  to
\begin{eqnarray}
A_i&=& 2 {\mathcal F} \left(\frac{-J_{ij} x_j+x_i x_{2n-1}}{(1+\vec{x}^2)^2}\right)\quad ,\quad i,j=1, 2, \dots, (2n-2) \nonumber\\
A_{2n-1}&=& {\mathcal F} \left(\frac{1-\vec{x}^2+2x_{2n-1}^2}{(1+\vec{x}^2)^2}\right).
\label{gauge-p}
\end{eqnarray}
When $n=2$ (i.e.,  a 3 dimensional gauge field $A_\mu$), this precisely reproduces 
the form of the original Loss-Yau gauge field \cite{loss}, 
although the coefficient ${\mathcal F}$ has not yet been determined.

We now turn to the stereographic projection of the Dirac spinor and operators in $S^{2n-1}$ from the ones of $\textbf{R}^{2n-1}$. The degeneracy of the fermionic zero modes will be deduced by group theoretic arguments for spinors in $S^{2n-1}$. We may define  $2^{n-1}\times 2^{n-1}$ Dirac matrices $\gamma_\mu$ ($\mu=1, \dots, (2n-1)$) for $\textbf{R}^{2n-1}$, and $2^{n}\times 2^{n}$ Dirac matrices $\Gamma_a$ ($a=1, \dots, 2n$) for $\textbf{R}^{2n}$, with those related as
\begin{eqnarray}
\Gamma_\mu =\begin{pmatrix}
{0 & i \gamma_\mu\cr
-i \gamma_\mu &0} 
\end{pmatrix}
\quad;\quad \Gamma_{2n} =\begin{pmatrix}
{0 & \textbf{1}\cr
\textbf{1} &0} 
\end{pmatrix}
\quad;\quad \Gamma_{2n+1} =\begin{pmatrix}
{\textbf{1} & 0\cr
0 & -\textbf{1}} 
\end{pmatrix}
\end{eqnarray}
Then the spin angular momentum operators in $\textbf{R}^{2n}$ are block-decomposed as
\begin{eqnarray}
\Sigma_{ab}\equiv \frac{1}{4i}[\Gamma_a, \Gamma_b] =\begin{pmatrix}
{\Sigma_{ab}^{+} &0\cr
0& \Sigma_{ab}^{-}}
\end{pmatrix} ,
\end{eqnarray}
where $\Sigma_{\mu\nu}^{\pm}=\sigma_{\mu\nu}\equiv \frac{1}{4i}[\gamma_\mu, \gamma_\nu]$ and $
\Sigma_{\mu, 2n}^{\pm}=\pm \frac{1}{2}\,\gamma_\mu$. By choosing the positive-helicity sector, we will only  
be using $\Sigma_{ab}^{+}$ in this work.
We also need the $2n$ dimensional orbital angular momentum generators 
\begin{eqnarray}
L_{ab}\equiv -i \left(y_a \frac{\partial}{\partial y_b}-y_b \frac{\partial}{\partial y_a}\right).
\label{orbital-2n}
\end{eqnarray}

A $2^{2n-1}$ component Dirac spinor $\psi$ in $(2n-1)$ dimensions is related 
to the $2^{2n-1}$ component Dirac spinor $\Psi_{+}$, which is the upper half of the 
full spinor (as obtained 
with the projection matrix  $(1 +\Gamma_{2n+1})/2$) through the following stereographic projection \cite{banerjee}:
\begin{equation}
\psi= \Omega^{-n+1/2} V^\dagger \Psi_{+}.
\label{dirac-proj}\end{equation}
In (\ref{dirac-proj}), the matrix
\begin{equation}
V = \frac{1}{\sqrt{2}}\left(\textbf{1}+i \gamma_\mu x_\mu\right)
\end{equation}
and   a simple factor $\Omega={1+\vec{x}^2}/{2}$ are related by
\begin{equation}
V^\dagger V=\Omega \textbf{1}.
\end{equation}
Then a stereographic projection of the {\it free} Dirac operator  is expressible as
\begin{eqnarray}
i\gamma_\mu \frac{\partial}{\partial x_\mu}\psi=\Omega^{-n+1/2}V^\dagger \left[\Sigma_{ab}^+ L_{ab}+\left(n-\frac{1}{2}\right)\textbf{1} \right]\Psi_{+}.
\label{free-dirac}
\end{eqnarray}
We now observe that for the $2n$ dimensional gauge field ${\mathcal A}_a$ defined in (\ref{sd-gauge}) and the $(2n-1)$-dimensional gauge field $A_\mu$ defined in (\ref{gauge-p}), this projection property of the free Dirac equation will be maintained given the inclusion of the gauge field interaction term
\begin{equation}
\gamma_\mu A_\mu \psi =\Omega^{-n+1/2}V^\dagger\Sigma_{ab}^+ (y_a {\cal A}_b-y_a {\cal A}_b)\Psi_{+}.
\label{gauge-inter}
\end{equation}
Then the zero-mode equation in $\textbf{R}^{2n-1}$,
\begin{equation}
\gamma_\mu (\partial_\mu -i A_\mu)\psi=0 \label{3zeroeq}
\end{equation}
can be lifted to a zero-mode equation on $S^{2n-1}$:
\begin{equation}
\left[\Sigma_{ab}\left(L_{ab}+y_a {\cal A}_b-y_b {\cal A}_a\right)+(n-1/2)\textbf{1}\right]\Psi_{+}=0. \label{4zeroeq}
\end{equation}
The solutions of this Dirac equation may be written in terms of the spinor spherical harmonics in $\textbf{R}^{2n}$.

\section{Zero modes in 3D}\label{sec3d}
In \cite{dunne}, this last equation was analyzed  in detail when $n=2$. The $4$-dimensional gauge field can 
be written as ${\mathcal A}_a=-({\mathcal F}/2) \bar{\eta}_{ab}^3 y_b$, where $ \bar{\eta}_{ab}^3$ is 
the 3rd isospin component of the standard $4$D 't Hooft tensor \cite{thooft,jackiwrebbi}. 
Then the zero-mode equation (\ref{4zeroeq}) becomes 
\begin{eqnarray}
\left[\Sigma_{ab}^{+}({L}_{ab}+2y_a{\cal A}_b)+3/2\right] \Psi_{+}=\left(4\vec{S}\cdot\vec{L}+3/2 -{\mathcal F}\sigma_3 /2 \right)\Psi_{+}=0. \label{dirac3D}
\end{eqnarray}
Here, $\vec{S}$ and $\vec{L}$ are angular momentum operators of spin $1/2$ and $l$ (= half integer), respectively. Note that we have chosen the half-integral representation of orbital angular momentum following the usual convention for $SO(4)$. 
The total angular momentum is $\vec{J}=\vec{S}+\vec{L}$,  and its eigenstates are classified by the spinor spherical harmonics\cite{bogomolny}
\begin{eqnarray}
\begin{pmatrix}
{\pm \sqrt{l+1/2\pm M} \,Y^{l}_{m, M-1/2}\cr
\sqrt{l+1/2\mp M}\, Y^{l}_{m,M+1/2}}
\end{pmatrix},  \label{CGspinor}
\end{eqnarray}
with  $-j\leq M\leq j$ and $-l\leq m\leq l$ for $j=l\pm 1/2$. 
The numerical factors in front of the four dimensional spherical harmonics $Y^l_{m, M\mp1/2}$ are the Clebsh-Gordan coefficients. 
The spin-orbital part $4\vec{S}\cdot\vec{L}$ has a value $2l$ when $j=l+1/2$ and $-2l-2$ when $j=l-1/2$. 
When ${\cal F}=4l+3$ for the case $j=l+1/2 $ and $M=l+1/2$, we get the fermion zero modes
\begin{eqnarray}
\Psi_{+}= Y^l_{m, l}    \begin{pmatrix}
{1\cr
 0}.
\end{pmatrix} \label{3dzero}
\end{eqnarray}
Here, $m$ has any  value in $-l,\cdots, l$ and so there are $2l+1$ zero modes.  
Therefore, we have $L+1$ (now $L=1,2,\cdots$) zero modes when  ${\cal F}=2L+3$. 
Note that any of them can be expressed as the product of a simple spinor $(1,0)^{T}$ 
and an orbital function denoted by  $Y^l_{m, l}$.

As noticed in \cite{adam2}, these zero modes and the gauges fields in 3 dimensions are closely related with the Hopf map from $S^3$ to ${\bf R}^2$. 
The basic Hopf map is defined by the complex valued function
\begin{equation}
\chi_H=\frac{2(x_1+i x_2)}{2x_3 -i (1-\vec{x}^2)} \equiv {\cal S} e^{i\sigma}
\end{equation} 
with a modulus ${\cal S}$ and a phase $\sigma$.
Using the variables $y_a$ in (\ref{stereo}), we can write this map as
\begin{equation}
\chi_H=\frac{y_1+i y_2}{y_3 -i y_4}. \label{hopf}
\end{equation}
One may easily identify that ${\cal S}^2=(y_1^2+y_2^2)/(y_3^2+y_4^2)$ and 
$\sigma=\tan^{-1}(y_2/y_1)+\tan^{-1}(y_4/y_3)$. 
On the other hand, the spherical harmonic function $Y^l_{m l}$ has the form
\begin{equation}
Y^l_{m l}\propto (y_3-iy_4)^{l+m}(y_1+i y_2)^{l-m}. \label{4dY}
\end{equation}
Given the Hopf map $\chi_H$ in (\ref{hopf}), this spherical harmonic function becomes
\begin{eqnarray}
Y^l_{m l} &\propto & \left( \frac{y_1+i y_2}{y_3-i y_4} \right)^{l-m}(y_3^2+y_4^2)^l \\
&=&e^{i L (\varphi-\sigma)} \frac{\chi_H^{L-n}}{(1+{\cal S}^2)^{L/2}},
\end{eqnarray}
with $L=2l$ and $n=m+L/2$ ($n=0,1,\cdots,L$).  Inserting this into (\ref{3dzero}),  one obtains 
all of the zero mode solutions in Eq(20) of \cite{adam2} up to  overall normalization constants. 

There have been various efforts to understand the topological nature 
of these zero modes in three dimensions.
 Erd\"os and Solovej \cite{erdos}  gave an elegant interpretation of these zero-mode-supporting 
 gauge fields  in terms of pull-backs (to ${\bf R}^3$) of 2 dimensional magnetic fields. 
Further results have been found in \cite{adam,elton}. 
There is a work in which these Abelian gauge fields have been understood in terms of projections of 
SU(2) gauge fields \cite{jackiw-pi}.

\section{Symmetry of the Dirac operator: SU($n$)$\times$U($1$)}
We now turn to the case of general odd ($2n-1$) dimensions. First note that the free Dirac operator (\ref{free-dirac}) on the sphere $S^{2n-1}$ is invariant under any transformation of SO($2n$) which is the rotation group of $\textbf{R}^{2n}$. 
The wave function $\Psi_{+}$ can then be classified by representations of SO($2n$). In Dynkin's notation,
a representation of SO($2n$) is denoted by $[l_1,l_2, \cdots, l_n]$. 
The Casimir invariant of this representation is given by \cite{wybourne}
\begin{equation}
C_2([l_1,l_2, \cdots, l_n])= \sum_{i=1}^{n}l_i(l_i+2n-2i).
\label{casimir}
\end{equation}
The orbital part can be expressed by the spherical harmonics in $2n$ dimensions and it belongs to the representation $[L,0,\cdots,0]$. The Casimir invariant for this representation has the value
$C_2(L)=L(L+2n-2)$ with integer $L$ ($=0,1,2, \cdots$), and the dimension of it is 
\begin{equation}
N_d([L,0,\cdots,0])= 2(L+n-1)\frac{(L+2n-3)!}{L!(2n-2)!}.  \label{so2ndim}
\end{equation}
There are two fundamental spinor representations in SO($2n$). Each of them has a definite helicity.
In this work, we have chosen a positive one and denote it by $[0,\cdots,0,1]$.  It has a Casimir 
invariant with the value $n(2n-1)/4$ and the dimension
\begin{equation}
N_d([0,\cdots, 0,1])= 2^{n-1}.
\end{equation}
A direct product of these two representation $[L,0,\cdots,0]$ and $[0,\cdots,0,1]$  is decomposed into two irreducible representations $J_{+}=[L,0,\cdots,0,1]$ and $J_{-}=[L-1,0,\cdots,0,1]$. Then the Casimir invariant of the total angular momentum  has 
a value $C_2(J_{+})=L(L+2n-1)+n(2n-1)/4$ or
$C_2(J_{-})=(L-1)(L+2n-2)+n(2n-1)/4$ for each of the two different representations.
The spin-orbit interaction term $\Sigma_{ab}^{+} L_{ab}$ can be expressed in terms of these Casimir invariants. 
It is automatically diagonalized and has the value
\begin{eqnarray}
\Sigma_{ab}^{+}L_{ab}=C_2(J)-C_2(L)-C_2(S)= 
\left\{\begin{matrix}
{L\cr
-(L+2n-2)}
\end{matrix}\right.
\end{eqnarray}
for each case. One may identify the fermion zero modes after expressing all 
the spinor spherical harmonics in explicit forms as we did in the previous section. 
However, it is a massive job to find all the related Clebsh-Gordan Coefficients even though all of the $2n$-dimensional spherical harmonics are known \cite{barut}. 

Here, we introduce an alternative and direct way using the symmetry of the Dirac equation.
The free Dirac operator has an SO($2n$) symmetry.
The presence of the gauge field in (\ref{4zeroeq}) breaks this SO($2n$) symmetry. 
However, it is possible to show that SU($n$)$\times$U(1), which is
a subgroup of SO($2n$), is a symmetry group of the Dirac equation(\ref{4zeroeq}).
In order to see this, let us introduce the following $2n$ matrices\cite{georgi}:
\begin{eqnarray}
b_i&=& \frac{1}{2}(\Gamma_{2i-1} -i \Gamma_{2i})   \nonumber   \\
b_i^\dagger &=& \frac{1}{2}(\Gamma_{2i-1} +i \Gamma_{2i}) \label{bgamma}
\end{eqnarray}
for $ i=1,2,\cdots, n$. 
These will then satify the following commutation relations:
\begin{eqnarray}
\{ b_i, b_j \}= \{ b_i^\dagger , b_i^\dagger \}=0, \quad \{ b_i , b_j^\dagger \}=\delta_{ij} . \label{comm-rel}
\end{eqnarray}
We may construct SU($n$) generators in the spin space as
\begin{equation}
T_S^\alpha= b_i^\dagger [T_\alpha]_{ij} b_j
\end{equation}
using the matrix elements $ [T_\alpha]_{ij}$ of the defining representation of SU($n$).
Note that the $T^\alpha$'s are expressible as a linear combination of a part of the SO($2n$)
generators
\begin{equation}
T_S^\alpha=  [T_\alpha]_{ij} \left( \frac{i}{2}\Sigma_{2i-1,2j-1}+\frac{1}{2}\Sigma_{2i-1,2j}
-\frac{1}{2}\Sigma_{2i,2j-1} + \frac{i}{2}\Sigma_{2i,2j} \right) 
\label{subalgebra}
\end{equation}
and generate a closed subalgebra. One may easily verify that the above generators 
satisfy the SU($n$) algebra. By replacing the spin generators
 $\Sigma_{a\,b} $  in (\ref{subalgebra}) with the orbital angular momentum generators $L_{a\,b}$,
 we can construct generators $T_L^\alpha$ acting on the coordinates $y_a$ as
\begin{equation}
T_L^\alpha=  [T_\alpha]_{ij} \left( \frac{i}{2}L_{2i-1,2j-1}+\frac{1}{2}L_{2i-1,2j}
-\frac{1}{2}L_{2i,2j-1} + \frac{i}{2}L_{2i,2j} \right).
\label{subalgebra2}
\end{equation}
It is convenient to define a complex variable $z_i$ and its complex conjugation $\bar{z}_i$ 
in therms of the  pair of  coordinates $y_{2i-1}$ and $y_{2i}$, so that
\begin{equation} 
z_i =y_{2i-1} +i y_{2i} , \quad
\bar{z}_i= y_{2i-1} -i y_{2i},
\end{equation}
for $i=1,2,\cdots, n $. We then find the following commutation relations for the generators and $b_i$, $b_i^\dagger$, $z_i$ and $\bar{z}_i$:
\begin{eqnarray}
[ T_S^\alpha , b_i^\dagger ] &=& b_j^\dagger [T_\alpha]_{ji} \;, 
\qquad [T_S^\alpha, b_i] = - [T_\alpha ]_{ij} b_j ,\\ \label{trans1}
 [T_L^\alpha, \bar{z}_i ] &=& - [T_\alpha]_{ij} \bar{z}_j \; ,  
  \qquad [T_L^\alpha, z_i] = z_j  [T_\alpha ]_{ji} .
\label{trans}
\end{eqnarray}
These transform according to the defining representation of SU($n$).
Note that a set of generators, $L_{2k-1,2k}$ (or $\Sigma_{2k-1,2k}$ in the spinor space), with
$k=1,\cdots,n$, 
forms the Cartan subalgebra of SO($2n$). Then the sum of these generators 
\begin{equation}H_L=\sum_{k=1}^n L_{2k-1,2k} \label{u1}
\end{equation}
commutes with all generators of SU($n$) to make an Abelian subgroup, U(1).

Note that $ 2 \Sigma_{ab}(y_a {\cal A}_b)$ in (\ref{4zeroeq}) is the term breaking the SO($2n$) symmetry.
 It can be now written as
\begin{equation}
 \Sigma_{ab} y_a {\cal A}_b= -\frac{i}{4} [ \Gamma_a y_a , \Gamma_b {\cal A}_b ] . \label{broken}
\end{equation}
In this expression, $\Gamma_a y_a$ is invariant under an SO($2n$) 
transformation but $\Gamma_a {\cal A}_a=-\Gamma_a C_{ab}y_b{\cal F}/2$ is not. 
However, we can cast the second part into the following SU($n$) invariant form:
\begin{eqnarray}
\Gamma_a C_{a b} y_b&=&\Gamma_{2i-1} y_{2i} - \Gamma_{2i} y_{2i-1} \nonumber \\
&=&-i (b_i z_i -b_i^\dagger \bar{z}_i).  \label{osc}
\end{eqnarray}

\section{Zero modes in ($2n-1$) dimension}
\subsection{Number of Zero modes}
The Dirac equation (\ref{4zeroeq}) can be now written, in a manifestly SU($n$) invariant form, as
\begin{eqnarray}
\left( \Sigma_{ab}^{+} L_{ab} + (n-\frac{1}{2}) - \frac{1}{2}{\cal F} +{\cal F} b_i^\dagger b_j \bar{z}_i z_j \right)\Psi_{+} =0.
\label{invariant}
\end{eqnarray}
Apparently this is also invariant under the U(1) transformation generated by (\ref{u1}).
Remember that the first term in (\ref{invariant}) can be replaced with the value $L$ when the Dirac spinor 
$\Psi_{+}$ belongs to the representation $[L+1,0,\cdots,0,1]$.
When the magnitude of field strength ${\cal F}$ has the value
\begin{equation}
{\cal F}= 2L+2n-1,  \label{Fvalue}
\end{equation}
the fermionic zero modes can be obtained by imposing the condition 
\begin{equation}
b_j\Psi_{+}=0 \label{spin-cond} 
\end{equation}
for all $j$. This implies that the spinor part of
$\Psi_{+}$ should be a Clifford vacuum and  it is thus a singlet of SU($n$). 
To realize this vacuum in a simple form,
let us take the representation of the gamma matrices in (\ref{bgamma}), so that 
\begin{equation}
b_i= (\textbf{1}\otimes)^{i}\begin{pmatrix}
{0 & 0\cr 1 & 0}\end{pmatrix}(\otimes \sigma_3)^{n-i}
\end{equation} 
as in \cite{georgi}.
Then the spinor part of the zero modes in (\ref{invariant}) is 
uniquely determined by the Clifford vacuum in a simple form:
\begin{equation}
\begin{pmatrix}
{0\cr
1}\end{pmatrix} \otimes \cdots  \otimes
\begin{pmatrix}
{0\cr
1}\end{pmatrix}. \label{vacuum}
\end{equation}

Since the spinor part is a singlet, it becomes trivial to construct a product of the orbital and spinor parts.  
The orbital part of $\Psi_{+}$ can be found by breaking the $[L,0,\cdots,0]$ representation of SO($2n$) into  irreducible representations of SU($n$). The branching of the representation $[L,0,\cdots,0]$ in SO($2n$) then
becomes
\begin{equation}
[L,0,\cdots,0]_{SO(2n)} = \sum_{m=0}^L [L-m, 0, \cdots, 0, m]_{SU(n)} . \label{branch}
\end{equation}
(This is derived from the observation that the fundamental representation $[1,0,\cdots,0]_{SO(2n)}$ corresponds to $[1,0,\cdots,0]\oplus[0,0,\cdots,0,1]$ in SU($n$) \cite{slansky}). 
The quantity in the left side of this equation denotes a representation in SO($2n$) 
and the objects in the right 
side are representations in the subgroup SU($n$). One can verify this relation
by matching the dimension of the representations in both sides:
\begin{equation}
2(L+n-1) \frac{(L+2n-3)!}{L! (2n-2)!}=\sum_{m=0}^{L}\frac{(L+n-1)}{(n-1)}\frac{(L-m+n-2)!}{(n-2)!(L-m)!} \frac{(m+n-2)!}{(n-2)! m!} \label{dimSO}
\end{equation}
The representation with $m=0$, which is denoted by $[L,0,\cdots,0]$ (in the SU($n$) group),
 carries the desired orbital angular momentum $L$ and  has the dimension
\begin{equation}
N_d([L,0,\cdots,0])_{SU(n)}=\frac{(L+n-1)!}{(n-1)! L!}. \label{dim}
\end{equation}
A product of the orbital part denoted by $[L,0,\cdots,0]$  and the vacuum spinor in (\ref{vacuum}) gives us the zero modes. 
Hence, when (\ref{Fvalue}) holds, the formula in (\ref{dim}) determines the number of zero modes of the Dirac operator.

\subsection{Explicit forms of zero modes}
In this section,  we obtain explicit forms of the fermion zero modes. Since the representation $[L,0,\cdots,0]$ corresponds to the spherical harmonics on the $S^{2n-1}$ sphere, 
let us change the coordinates $(y_1,y_2,\cdots,y_n)$ into the angular variables: 
\begin{eqnarray}
y_1&=&\cos\phi_1 \prod_{i=1}^{n}\sin\theta_i ,  \qquad \qquad \quad y_2= \sin\phi_1 \prod_{i=1}^{n}\sin\theta_i  \\
y_{2k-1}&=&\cos\phi_k \cos\theta_k \prod_{i=k+1}^{n}\sin\theta_i, \quad
y_{2k}  =\sin\phi_k \cos\theta_k \prod_{i=k+1}^{n}\sin\theta_i, \quad ( k=2,\cdots n) 
 \label{n-angle}
\end{eqnarray}
where each $\phi_i$ ($i=1,2\cdots,n$) takes a value in $[0,2\pi]$ and $\theta_i$ ($i=2\cdots,n$) takes a value in $[0,\pi/2]$. 
With these angles, the spherical harmonics in $2n$ dimensions have the form \cite{barut}
\begin{eqnarray}
Y^{L\,(l_2,\cdots,l_{n-1})}_{m_1,\cdots,m_n}=N_n^{-1/2}\prod_{k=2}^n \sin^{2-k}\theta_k  d^{J_k}_{M_k\,M_k'}(2\theta_k)\exp(i \sum_{j=1}^n m_j \phi_j) \label{Ylm}
\end{eqnarray}
where ${m_1,\cdots,m_n}$ denotes the components of a  weight vector belonging to the  $[L,0,\cdots,0]$ representation and the sub-angular-momentum numbers $l_n\equiv L\ge l_{n-1}\ge\cdots\ge l_2\ge 0$ with $l_1\equiv m_1$ have been introduced. 
These satisfy the relations
\begin{equation}
(|m_k|+|l_{k-1}|) =  l_k -2s_k, \quad s_k=0,1,\cdots [l_k/2], \quad k=2,\cdots,n . \label{constraints}
\end{equation}
In (\ref{Ylm}), the $d^J_{M M'}$'s  are the Wigner $d$-functions of the ordinary rotation group SO(3). 
The quantum numbers of those functions are related with the above numbers as
\begin{eqnarray}
J_k&=& \frac{1}{2} (l_k+k-2), \quad M_k=\frac{1}{2}(m_k-l_{k-1} -k +2), \nonumber \\
   M_k'&=&\frac{1}{2}(m_k-l_{k-1} -k +2),   \quad  \quad(k=2,\cdots,n) .
\end{eqnarray}
Then each $M_k$ or $M_k'$ has one of $2J_k+1$ values between $[-J_k, J_k]$. Combining these restrictions with the relations in (\ref{constraints}), one should easily recognize that there are  $(l_{k}-l_{k-1}+1$ possibilities for each $d^{J_k}_{M_k M_k'}$ when $k=n,n-1,\cdots,3$ 
and  $(2l_2+1)^2$ possibilities when $k=2$. 
Therefore,  we  count the total number of possibilities in the spherical harmonics (\ref{Ylm}) to be
\begin{eqnarray}
\sum_{l_{n-1}=0}^L(L-l_{n-1}+1)\sum_{l_{n-2}=0}^{l_{n-1}}(l_{n-1}-l_{n-2}+1)\cdots \sum_{l_2=0}^{l_3}(l_3-l_2+1)(l_2+1)^2. \label{dimY}
\end{eqnarray}
This sum reproduces the dimension of the representation in the left side of (\ref{so2ndim}). 

The branching rule in (\ref{branch}) says that a suitable choice of the above spherical harmonics forms a desired representation in SU($n$). For this purpose, let us first note that the generators in the Cartan subalgebra of SO($2n$) can be represented by the differential operators of the following simple form:
\begin{equation}
L_{2k-1,2k}=\frac{1}{i} \frac{\partial}{\partial\phi_k} .
\end{equation}
Then the generator of the U($1$) group has the form $H_L=\sum_{k=1}^n (-i) {\partial}/{\partial\phi_k}$.
We also recall that SU($n$)$\times$U(1) is the symmetry group of the Dirac operator and that the zero modes are classified by the quantum numbers for a representation of the group.

The spherical harmonics in (\ref{Ylm}) are eigenfunctions of the U(1) generator 
and  are classified by the eigenvalue $\sum_{k=1}^{n} m_k$.  By fixing this eigenvalue to be $L$, one may have a specific SU($n$)$\times$U($1$) representation. 
Therefore,  the spherical harmonics in (\ref{Ylm}) satisfying the constraint 
\begin{equation}
\sum_{k=1}^{n} m_k=L \label{constraint}
\end{equation}
form
the desired $[L,0,\cdots,0]$ representation in SU($n$). The dimension of this can be counted in a  
similar manner as done above. 
We have the same number of possibilities $ (l_{k}-l_{k-2}+1)$ when $k=n,n-1,\cdots,3$. But, when $k=2$,  $M_2'=(m_2+l_1)/2=(m_2+m_1)/2$ is now fixed as $M_2'=(L-\sum_{k=3}^n m_k)/2$ because of the constraint (\ref{constraint}) and there are $l_2+1$ possibilities of $M_2$ instead of $(l_2+1)^2$. 
The total number of possibilities is now given by
\begin{equation}
\sum_{l_{n-1}=0}^L(L-l_{n-1}+1)\sum_{l_{n-2}=0}^{l_{n-1}}(l_{n-1}-l_{n-2}+1)\cdots \sum_{l_2=0}^{l_3}(l_3-l_2+1)(l_2+1).
\end{equation}
This sum reproduces the result in (\ref{dim}).  Using the spherical harmonics, one determine the fermion zero modes in ($2n-1$) dimension as
\begin{eqnarray}
\psi&=&\frac{1}{\sqrt{2}}\left(\frac{1+\vec{x}^2}{2}\right)^{-n+1/2}\left(1+i\gamma_\mu x_\mu\right) \Psi_{+} \nonumber \\
\Psi_{+}&=& Y^{L\,(l_2,\cdots,l_{n-1})}_{m_1,\cdots,m_n} 
\begin{pmatrix}{0\cr 1}\end{pmatrix} \otimes \cdots  \otimes
\begin{pmatrix}{0\cr 1}\end{pmatrix},
\end{eqnarray}
with the constraint (\ref{constraint}).

Let us make some comments on  two specific cases $n=2,3$. In 3 dimensions, the spherical harmonics $Y^{L/2}_{M,M'}$ with the condition $M=(m_1+m_2)/2=L/2$ are nothing but the harmonics in Sec.\ref{sec3d},
 and these become the functions in (\ref{4dY}) when the angle variables are converted into the coordinates $(y_1,y_2,y_3,y_4)$. 
In 5 dimensions (i.e., when $n=3$), the same kind of representation of the SU(3) group was studied 
a long time ago\cite{beg}, and exactly the same spherical harmonics were found.

\section{Conclusion}
We have studied the zero modes of the Abelian Dirac operator in odd dimensions. 
It turns out that the symmetry group of the Dirac operator is SU($n$)$\times$U($1$) when the dimension of the space is $2n-1$ ($n\ge 2$). We have determined the condition for the existence of such zero modes. The number of zero modes is determined by the dimension of a suitable representation of the symmetry group. We have also found the explicit forms of the zero modes using the spherical harmonics 
in terms of angular variables for the $2n$ dimensional space.

Some physical quantities have  integral values, depending on the dimension of the space $2n$ and the orbital angular momentum $L$.   Note that we have zero modes only when the magnitude of the field strength is an odd integer ${\cal F}=2L+2n-1$.  The number of zero modes is also an integer ${(L+n-1)!}/{(n-1)! L!}$.
In three dimensions, there have been some efforts to understand 
these integers on the basis of the Hopf index\cite{adam, erdos}. In any odd dimension, the Chern-Simons number could be a good object for such a consideration.
However,  it is proportional to ${\cal F}^n=(2L+2n-1)^n$ and has a structure 
differing from the number of zero modes.
To find a mathematical understanding among these integers should be an interesting problem in mathematical physics.

\acknowledgments
The author thanks G. Dunne, Jae-Hoon Kwon and A. Medved for helpful discussions. This work was supported in part by the University of Seoul 2007 Research Fund.

\end{document}